# Differential Learning for Robust Prediction of Thermal Stability with Application to Energetic Materials

*Megan C. Davis*[†*], *R. Seaton Ullberg*[†*], *Jeremy N. Schroeder*[†‡], *Andrew H. Salij* [†], *Marc J. Cawkwell*[†], *Christopher J. Snyder*[¶], *Ivana Matanovic*[†], *and Wilton J. M. Kort-Kamp*[†]

† Theoretical Division, Los Alamos National Laboratory, Los Alamos, NM 87545, United States

‡ Department of Mechanical and Aerospace Engineering, Texas Tech University, Lubbock, TX 79409, United States

¶ Weapon Stockpile Modernization Division, Los Alamos National Laboratory, Los Alamos, NM 87545, United States

Corresponding authors: megand@lanl.gov, sullberg@lanl.gov, kortkamp@lanl.gov

[*] These authors contributed equally.

## ABSTRACT

Predicting thermal stability during handling and storage is essential for the design of safe and reliable energetic materials. However, experimental measurements vary significantly across laboratories due to differences in protocols and analysis methods, making it difficult to train reliable predictive models. We address this challenge through differential learning. Rather than predicting absolute decomposition temperatures, we instead train message passing neural networks to predict relative differences between pairs of molecules. This approach reduces sensitivity to systematic experimental errors and achieves >85% accuracy in ranking compounds by thermal stability, outperforming conventional regression methods on the same heterogeneous dataset. To understand what drives these predictions, we compare neural network models with interpretable alternatives built from descriptors derived from ab initio calculations and cheminformatics software. This analysis identifies bond dissociation enthalpy as a key determinant of thermal stability rankings, providing further insight into the complex chemistry of thermal decomposition. The differential learning framework generalizes across model architectures, from graph neural networks to classical descriptor-based approaches. Our results demonstrate that learning relative properties rather than absolute values offers a practical solution for modeling noisy experimental data, with direct applications in materials design where thermal stability predictions inform safety protocols.

---

## 1. Introduction

Thermal decomposition kinetics determine the temperature-time limit at which organic molecules can be processed, stored, or deployed without undergoing irreversible chemical changes.[1] The decomposition temperature ($T_d$) at which molecular breakdown occurs is widely adopted as a proxy for thermal stability, despite its dependence on heating rate and exposure time, and serves as a key parameter in materials screening and qualification workflows.[2–6] Accurate prediction of this quantity would streamline materials discovery and development by reducing our reliance on costly and time-consuming molecule synthesis, purification, and thermal characterization. Experimentally, decomposition temperatures are commonly determined using differential scanning calorimetry[7] and related thermal analysis techniques,[8] but the reported values are sensitive to experimental protocols. Heating rate, sample mass, thermal contact, and the choice of reporting convention such as onset or peak temperature can all influence the reported decomposition temperature.[9] Beyond intrinsic measurement uncertainty, these factors introduce systematic, source-dependent variability into compiled datasets, while the experimental context required to interpret such differences is often incomplete or inconsistently reported. As materials databases increasingly aggregate values from heterogeneous literature sources, this variability becomes embedded as latent noise, fundamentally limiting the reliability of conventional data-driven modeling.

Machine learning (ML) has become a useful methodology in molecular and materials informatics, enabling data-driven prediction of chemical, thermodynamic, and biological properties directly from structure.[10–13] Recent works have emphasized the development of molecular representations for property prediction, including graph-based neural networks that operate directly on the two-dimensional molecular graph[14–17] as well as descriptor-based models derived from cheminformatics fingerprints and electronic structure calculations.[18–22] While most studies formulate property prediction as a direct regression of absolute values, alternative learning strategies based on relative or paired representations have begun to emerge. For example, paired molecular representations have been used to classify relative potency differences in drug discovery settings, demonstrating that comparative formulations can enhance predictive performance when learning structure–activity relationships.[23] Similarly, pairwise difference regression frameworks train models to predict property differences between data points rather than the properties themselves, improving generalization and uncertainty estimation across chemical tasks.[24] Related twin-network regression architectures likewise operate on differences between targets to reformulate conventional supervised learning objectives.[25] Although these approaches were not developed specifically to address heterogeneous experimental data, they illustrate a broader principle: learning relative differences between instances can provide an alternative inductive bias to absolute regression. Thermal decomposition offers a compelling context in which such a formulation may be particularly advantageous. Because decomposition temperatures are compiled from diverse experimental protocols, systematic offsets arising from measurement conditions are embedded within the data. In this setting, directly regressing absolute temperatures risks entangling protocol-dependent artifacts with intrinsic molecular trends. Learning relative stability relationships instead provides a principled route to preserve chemically meaningful ordering while attenuating global shifts introduced by heterogeneous experimental conditions.

Here, we introduce a differential learning framework for the robust prediction of thermal stability in the chemical space of organic molecules containing only CHNO. Rather than directly regressing absolute decomposition temperatures, our approach learns pairwise differences between compounds, enabling consistent ranking under heterogeneous measurement conditions. We implement this strategy using message passing neural networks (MPNNs) operating on molecular graphs, together with complementary differential models constructed from interpretable descriptors derived from cheminformatics software and density functional theory (DFT) calculations. This framework is particularly well suited for experimentally heterogeneous properties whose reported values embed systematic protocol-dependent variability.

We apply this framework to energetic organic molecules, a class of materials in which thermal stability is both technologically critical and intrinsically coupled to molecular design.[26,27] In these systems, decomposition temperature serves as a primary metric of safe handling, storage lifetime, and operational reliability.[28] Compounds with insufficient thermal stability may undergo premature decomposition at elevated temperatures, whereas excessive stabilization tends to be correlated with diminished energetic performance, necessitating a careful balance between energy density and thermal robustness.[29–31] This energy–stability trade-off has long motivated the use of physically grounded descriptors to rationalize decomposition behavior. In particular, the minimum bond dissociation enthalpy (BDE) or "trigger-linkage" has been widely employed as a mechanistically motivated proxy for thermal robustness.[3,4,32,33] By combining differential graph-based models with descriptor-based differential regressors, we evaluate not only ranking performance but also whether the learned representations recover chemically meaningful determinants of decomposition. Feature attribution analysis identifies differences in oxygen balance and bond dissociation enthalpy as top predictors of relative thermal stability, directly linking the differential learning formulation to established mechanistic concepts. Together, this work positions differential learning as a principled framework for modeling experimentally heterogeneous materials properties, with thermal decomposition in energetic molecules serving as a stringent and mechanistically informative case study.

## 2. Results and Discussion

### *2.1 Model Workflow*

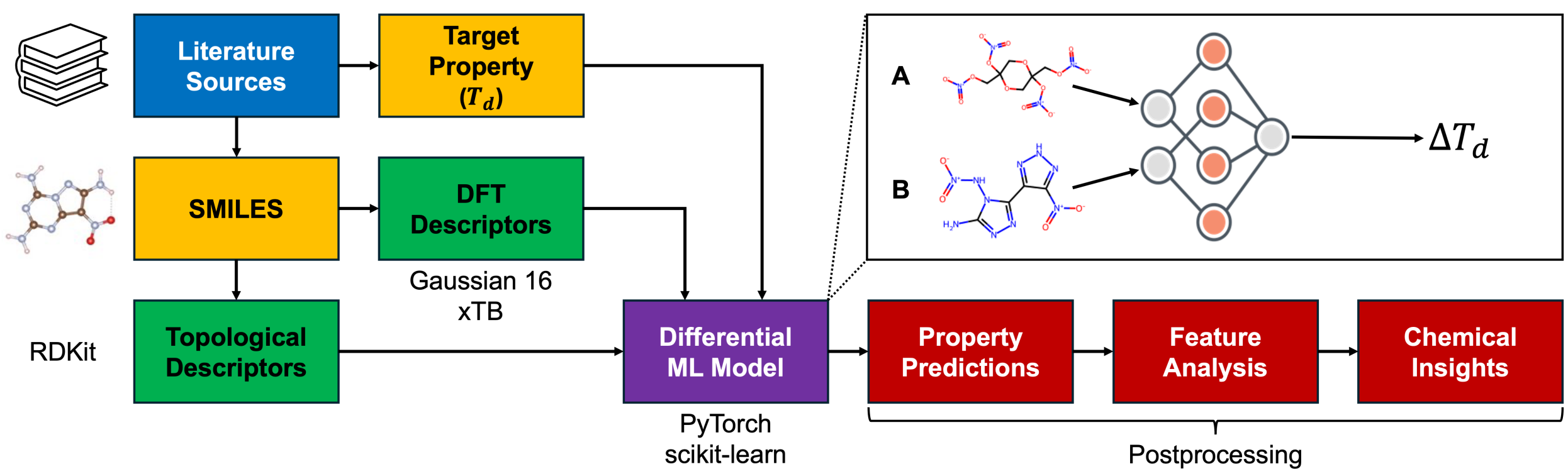

Figure 1 – High-throughput workflow used to collect data from literature, calculate descriptors, train machine learning models, and derive chemical insights. SMILES strings and associated $T_d$ values are collected from literature. A conformer search produces adequate starting geometry for DFT calculations while topological features are derived from the molecular graph. Chemprop and scikit-learn are used to train differential regression models to predict $\Delta T_d$ between pairs of molecules. Postprocessing results in SHAP scores that help to identify key descriptors in the prediction of thermal stability. The inset figure in the top right depicts a schematic of the differential regression model – two molecular graphs (or descriptor sets) are used as input and the $\Delta T_d$ between them is the targeted output. The sign of the $\Delta T_d$ prediction can be used to rank molecules by their thermal stability.

Figure 1 provides a schematic of our workflow for database curation, including the computation of properties from DFT (e.g., bond dissociation enthalpy) and subsequent model training to predict $\Delta T_d$. The full procedure is described in more detail in Section 4. We initially tested two approaches to train ML regression models to predict $T_d$ using our collected data: (i) interpretable ML models using DFT and cheminformatics descriptors as inputs, leveraging the scikit-learn package[34] and (ii) message passing neural networks implemented using Chemprop,[35] which are trained directly on molecular graphs. This direct regression approach exhibits underwhelming performance (Supplemental Information Figures S1 and S2), with the best model achieving a MAE of 48.9 °C and an $R^2$ of 0.70. Notably, predictive accuracy deteriorates at the upper and lower extremes of the $T_d$ distribution. This limitation is particularly consequential in materials discovery, where the objective is to identify candidates that optimize some target property and where high-performing outliers are therefore of greatest interest. A likely contributor to this poor predictive performance is the systematic noise present in our curated dataset. Because the dataset is compiled from measurements reported by multiple institutions, it is subject to systematic heterogeneity arising from differences in experimental protocols, reporting practices, and incomplete metadata availability. Moreover, the relatively small dataset size limits model generalizability and complicates the robust identification of underlying structure-property relationships. Collectively, these challenges motivate the development of alternative modeling strategies that are robust to data heterogeneity and sparsity while maintaining predictive fidelity.

To mitigate the effects of dataset noise, we reframed the task from predicting absolute $T_d$ values to learning the relative ordering of molecular performance using a differential learning framework. In this approach, the model receives a pair of molecules as input and predicts the difference in their $T_d$ values (Figure 1 inset). This formulation offers several advantages. First, constructing molecular pairs generates $\sim N^2$ training examples, substantially expanding the number of pairwise comparisons available during training. Second, because predictions are based on pairwise comparisons rather than absolute values, the model can focus on learning discriminative features that govern relative performance, improving robustness to systematic heterogeneity in the underlying data. Pairwise predictions can then be aggregated through a round-robin tournament to generate a global ranking without imposing arbitrary classification thresholds.

We evaluated both classification- and regression-based implementations of this differential learning strategy. In the classification formulation, the model predicts which molecule in a pair exhibits the higher $T_d$, whereas in the regression formulation it predicts the numerical difference in $T_d$ between the two molecules. The regression approach consistently outperformed classification, achieving an F1 score of 0.860 compared with 0.845 for the best classification model (Table S1). Consequently, the remainder of this work focuses on two complementary differential regression models: (i) a message-passing neural network (MPNN) operating on molecular graphs derived from SMILES representations, and (ii) a random forest (RF) model trained on differential molecular descriptors. Although the MPNN provides slightly superior predictive performance, both approaches offer complementary opportunities for model interpretation and the development of chemical insights (Section 2.3).

### *2.2 Model performance*

To assess the predictive performance of both the MPNN and RF models (implementation details outlined in Section 4.3), we apply them to the same held-out test set to collect consistent accuracy metrics. In Figure 2, the accuracy of each model is assessed as a function of the ground-truth $\Delta T_d$ between the pair of input molecules. Both the MPNN and RF models exhibit a trend of improved accuracy score as $\Delta T_d$ increases. Even at small separations, both models are always better than random choice (accuracy score > 0.5), which indicates that both architectures are effective in deriving structure from the data – either from the molecular graph in the MPNN model or from physical descriptors in the RF model. Both models converge to 100 % accuracy once $\Delta T_d$ surpasses approximately 250 °C. This result is intuitive, as we expect that either model should readily differentiate between molecules with drastically different $T_d$, but for molecules that are close in $T_d$, the decision boundary becomes narrower. This is despite the fact that the training data used to fit each model contains more pairs with a small $\Delta T_d$ as evidenced by the grey bars in Figure 2, which depict the number of pairs in each bin.

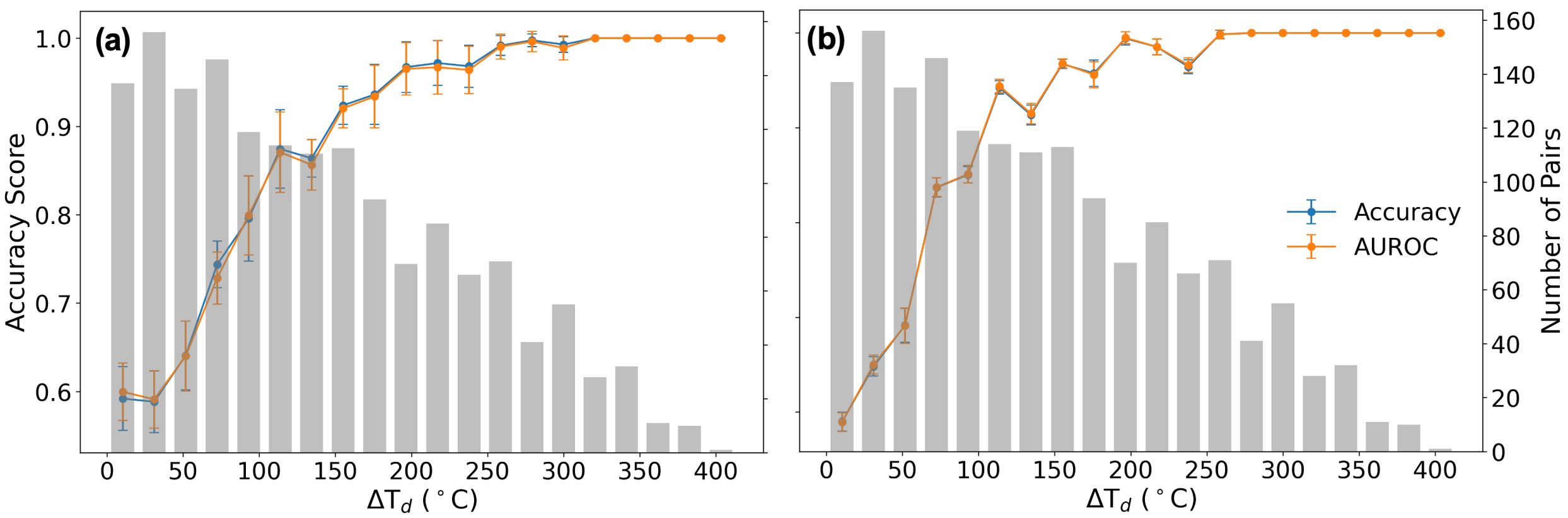

Figure 2 – Test set accuracy scores of the (a) MPNN model and (b) RF model in predicting which input molecule has a higher $T_d$ as a function of the difference in $T_d$ between the two molecules. The blue curve corresponds to accuracy, and the orange curve corresponds to the area under the receiver operating characteristic (AUROC). Error bars are computed as the standard deviation of each score within each $\Delta T_d$ bin. Grey bars indicate the number of molecule pairs in each $\Delta T_d$ bin.

Accurate ordinal ranking is needed for identifying new materials with enhanced thermal stability. To highlight this capability in both models, Figure 3 shows the parity between predicted rank order derived from a tournament-based ranking system (outlined in Section 4.4), and a rank order based on ground-truth $T_d$ for all molecules in the test set. Both the MPNN and RF models achieve a high pairwise accuracy, 87% and 86%, and a high $R^2$, 0.81 and 0.79, respectively. The points in Figure 3 are color-coded by the experimental $T_d$ of the molecule they represent to illustrate how experimental $T_d$ correlates with the rank predicted by the model. The position of 1,3,5-triamino-2,4,6-trinitrobenzene (TATB), a molecule often cited for its excellent thermal stability,[36] is represented by a vertical dashed line to serve as an example of how these surrogate models can be used during materials screening to identify new candidate molecules with high thermal stability. The nearly equivalent predictive performance exhibited by the two models further supports the idea that both architectures are capable of deriving structure from the data using different, but equally informative, molecular representations. In the following section, chemical and mechanistic insights into the factors governing thermal stability as predicted by these surrogates will be presented.

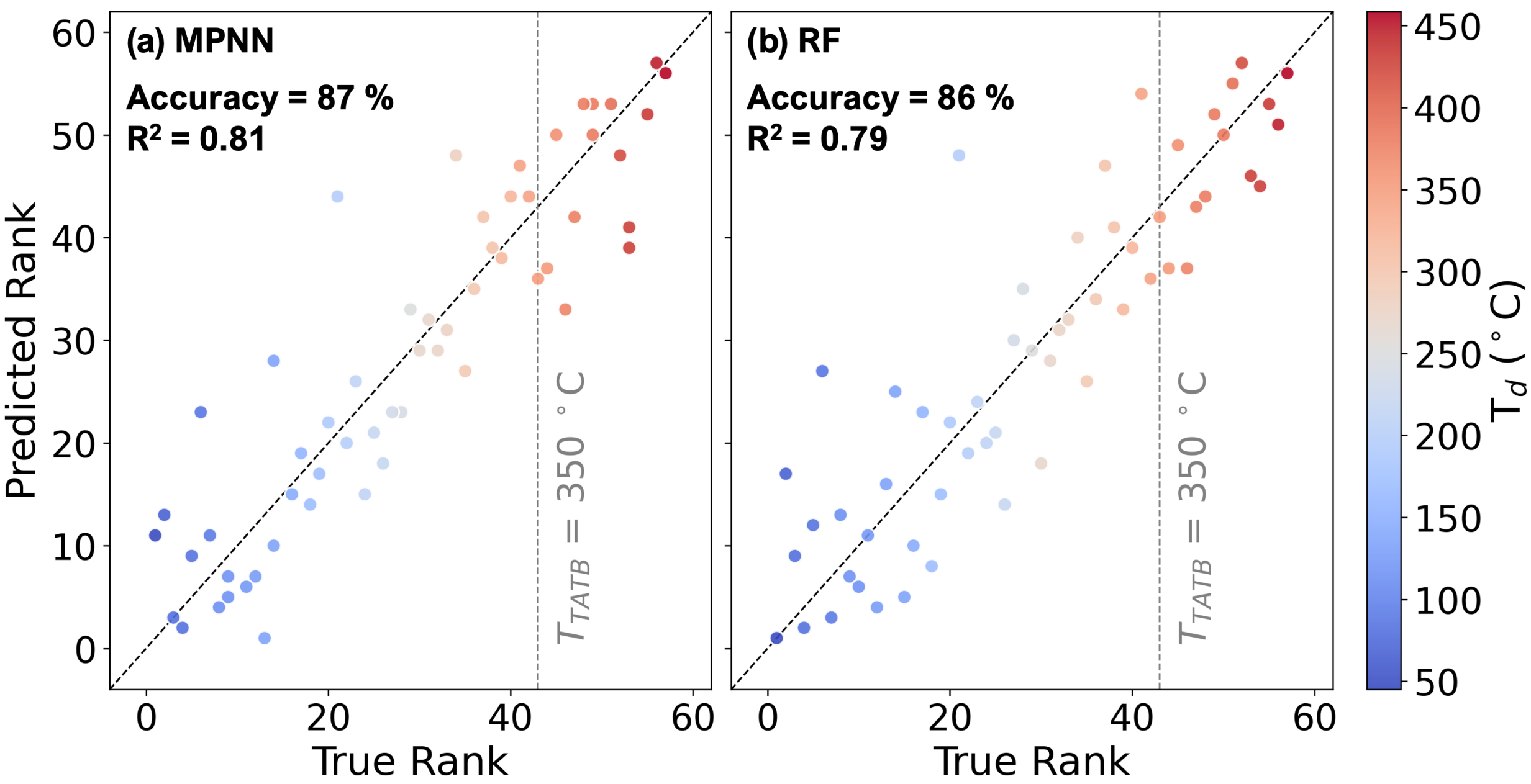

Figure 3 – Parity of (a) the MPNN model and (b) the RF model in predicting the order of each molecule in the test set ranked by its $T_d$. Points are colored by their experimental $T_d$ with red corresponding to higher values and blue corresponding to lower values as indicated by the color bar on the right. The position of TATB is indicated by a dashed line in both subplots to give a sense of the $T_d$ range which the test molecules cover.

*2.3 Model interpretability*

Beyond serving as surrogate predictors of thermal stability for unknown compounds, our MPNN and RF models can be interrogated using SHAP analysis (described in Section 4.3) to identify the descriptors that most strongly influence $\Delta T_d$ predictions, thereby providing interpretability and chemical insight into the factors governing thermal stability. Figure 4 presents the top descriptors identified by this analysis when applied to the RF model. Subplot (a) ranks the descriptors by their mean absolute SHAP score while subplots (b)-(g) plot the relationship between SHAP score and descriptor value for each of the top six descriptors, which indicates the directionality of the correlation. It is important to note that each descriptor is a differential descriptor, meaning that it captures the difference in the value of a property between molecules *A* and *B* rather than the magnitude of either one.

The difference in oxygen balance (%OB) is identified as the most important descriptor to predict $\Delta T_d$, with a mean SHAP value of 13.8 °C. This highlights a well-established inverse relationship between energetic performance and thermal stability in most energetic materials.[27] Molecules with an %OB near zero tend to be highly performant and are therefore less likely to be thermally stable.[31] From inspection of Figure 4b, it is clear that pairs of molecules with a large difference in %OB tend to produce the highest SHAP score for this descriptor indicating that this is the strongest proxy for difference in $T_d$ of the descriptors tested in this study.

Two descriptors that relate to bond dissociation also rank highly: the difference in mean BDE and the difference in minimum BDE with SHAP scores of 13.0 °C and 12.5 °C respectively. Referring to subplots (c) and (d) of Figure 4, it is evident that pairs of molecules with significant differences in their mean or minimum BDE are associated with the highest SHAP scores. This supports the argument often used in literature that the weakest bond in a molecule can act as a trigger which once broken could initiate the decomposition process. However, it is important to consider this descriptor in the full context of Figure 4a. Despite being some of the most important descriptors, they are clearly dominated by the combined contribution of the remaining 183 descriptors not explicitly listed. This suggests that the contribution of any individual feature is less important than the interaction or combination of them all which makes the use of BDEs alone (or any other individual descriptor for that matter) insufficient for screening thermally stable energetic materials. The remaining descriptors exhibit even less influence on the final $\Delta T_d$ prediction, further reinforcing the idea that it is the combination of these descriptors, rather than the value of any individual one, that enables accurate predictions by the RF model.

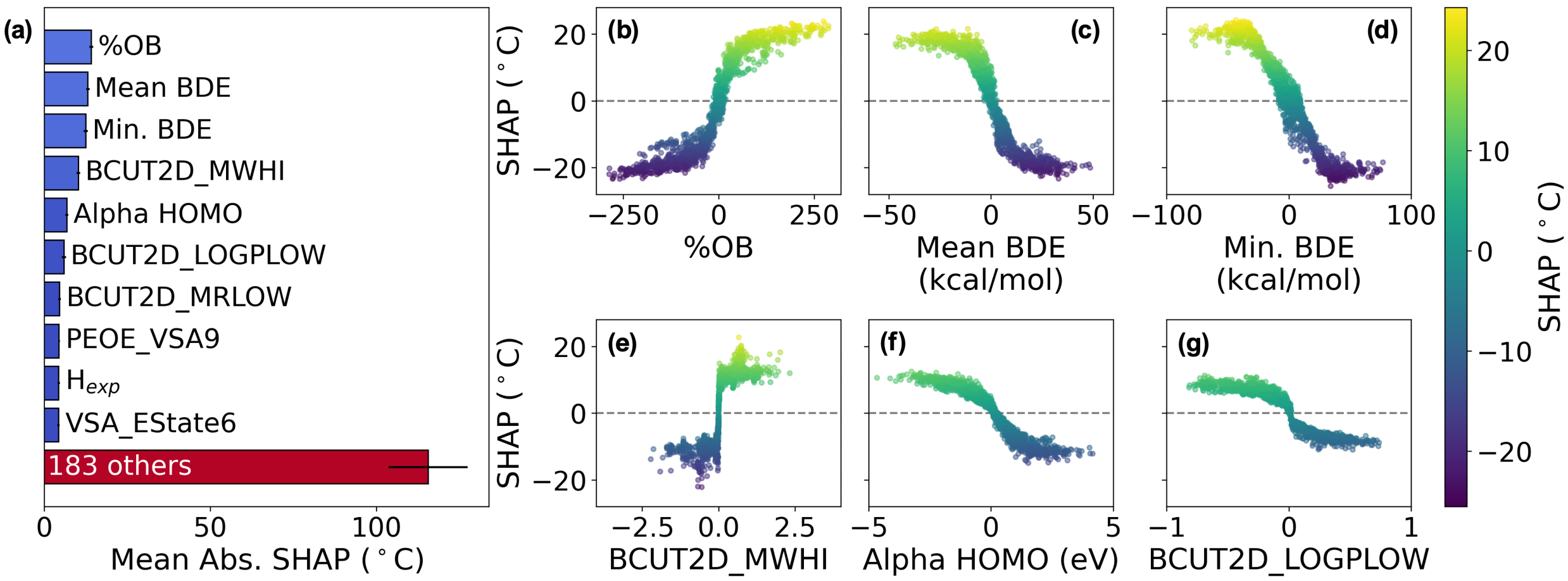

Figure 4 – SHAP importance scores of the top descriptors in predicting $\Delta T_d$ using the RF model. (a) Mean absolute SHAP score for the top ten most important descriptors alongside the sum of all remaining descriptors. (b-g) Scatter plots of SHAP score v. descriptor value for each molecule in the dataset. Each subplot corresponds to a different descriptor: %OB, Mean BDE, Minimum BDE, BCUT2D_MWHI, Alpha HOMO, and BCUT2D_LOGPLOW respectively. The colormap scales linearly with the Y-axis to emphasize the points with extreme values. Positive SHAP scores are associated with a positive contribution to predicted $\Delta T_d$ and negative SHAP scores are associated with a negative contribution to predicted $\Delta T_d$.

While Figure 4 illustrates the directional relationship of each descriptor with SHAP scores in the aggregate, it can also be informative to review results from the SHAP analysis on the individual molecule pair-level. Therefore, Figure 5 illustrates two representative scenarios encountered by the RF model, one in which molecule $A$ has a $T_d$ greater than molecule $B$ and vice versa. Figure 5a depicts a pair of molecules in which molecule $A$ is selected to be the molecule with the minimum $T_d$ in the dataset and molecule $B$ has a median $T_d$ value relative to the range of the whole dataset. Figure 5b reports the opposite scenario, where molecule $A$ has a median $T_d$ value and molecule $B$ is the molecule with the maximum $T_d$ in the dataset. As expected, each of the top descriptors identified in Figure 4 are present in both scenarios. However, it is important to note how specific differences in molecule composition and structure can impact the influence that a descriptor has – often changing the order from what was observed in the aggregate. For example, in Figure 5a, oxygen balance – the most important descriptor in aggregate – ranks low for the minimum/median pair of molecules since the two molecules do not have a substantial difference in their computed %OB (Δ %OB = 28.6). This means that the model must rely on other descriptors, namely mean BDE, to make a clear distinction between the two. Alternatively, in Figure 5b, the oxygen balance between the two molecules does differ significantly (Δ %OB = -200.3) and, in this case, can be used reliably to separate the pair.

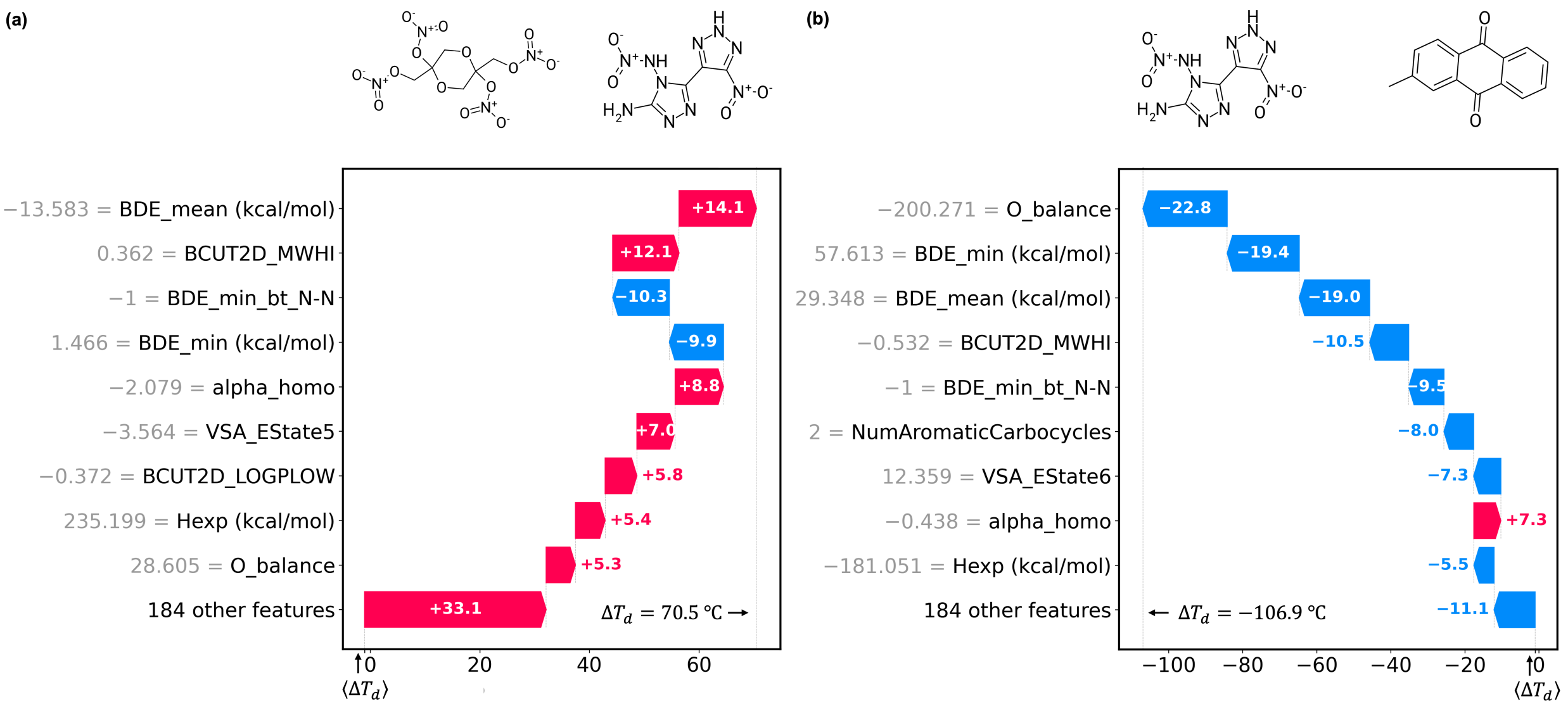

Figure 5 – Waterfall plots illustrating the SHAP analysis for two specific pairs of molecules. Subplot (a) describes how the RF model makes its prediction between the lowest $T_d$ (45 °C) molecule and a median $T_d$ (221 °C) molecule whereas subplot (b) illustrates the results for the median $T_d$ molecule and the highest $T_d$ (459 °C) one. These two scenarios compare how the RF model handles pairs with different orderings but similar $\Delta T_d$ magnitudes.

To better understand how BDEs affect thermal stability, Figure 6 illustrates the distribution of all BDE values grouped by bond type for the nine most prevalent bond types in our dataset.

Sorting the groups by their median BDE value, we observe a clear separation between the $NO_2$ linkages (C-$NO_2$, N-$NO_2$, and O-$NO_2$) and all other bond types. The weakness of R-$NO_2$ bonds (median BDE < 2 eV) relative to all other bond types reinforces a commonly used heuristic in the energetic materials literature that $NO_2$ groups, when present, often serve as the trigger linkage that initiates the decomposition reaction.[4,32,33,37,38] However, as shown in Figure 4a, this heuristic alone is insufficient to make an accurate prediction of thermal stability. Therefore, despite the prevalence of this idea in similar works focused on developing thermally stable energetic materials, we posit that the minimum or mean BDE of a molecule can be informative when used in conjunction with other topological and DFT-derived descriptors but should not be used as a direct proxy for thermal stability.

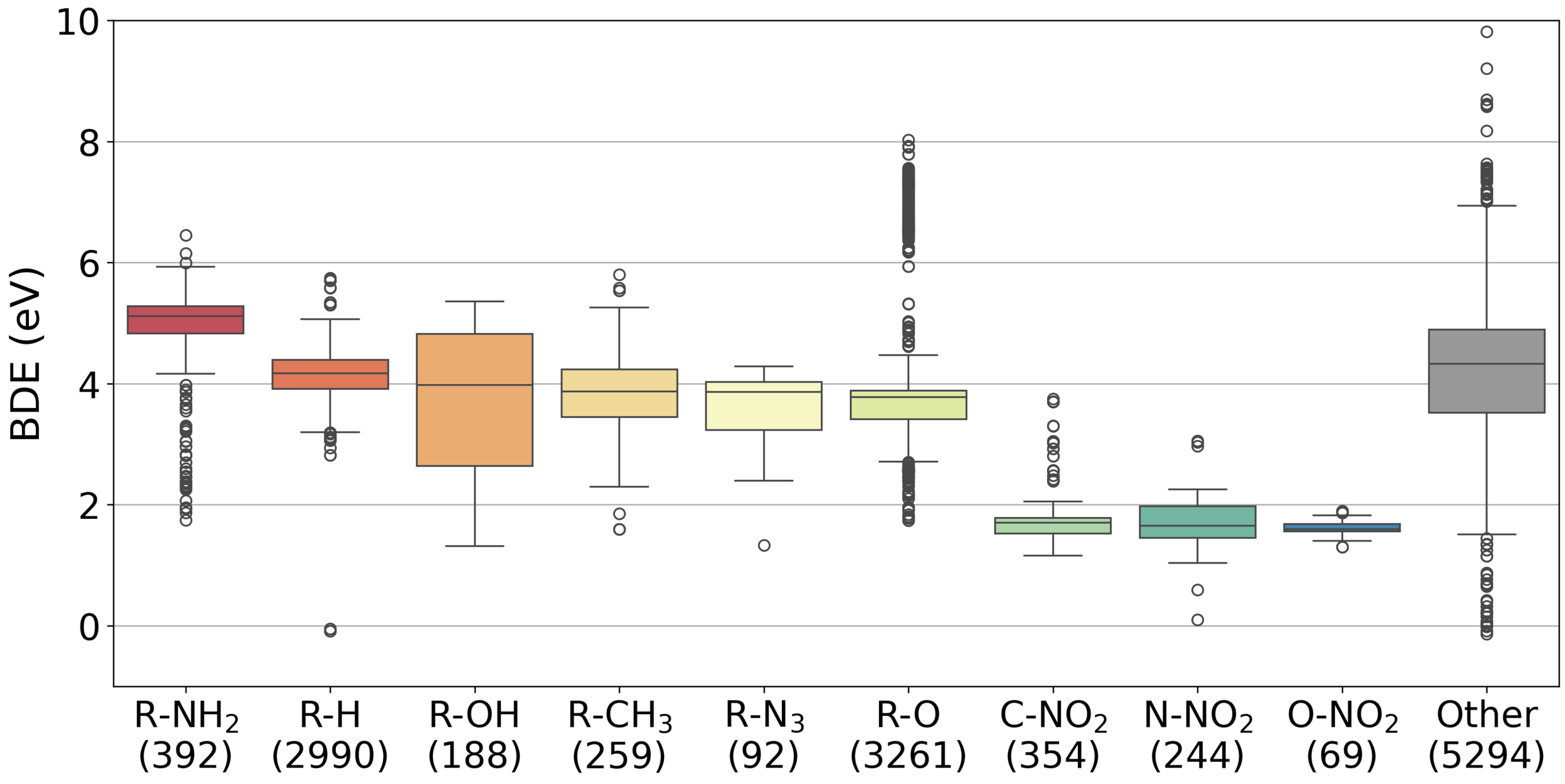

Figure 6 – Boxplot illustrating the distributions of BDE values for the most prevalent bond types in the dataset. The X-axis labels correspond to the type of bond being broken and the number in parentheses underneath it is the total number of bonds of that type. The bond types are ordered by their median BDE value with the highest starting at the left. All bond types that are not explicitly labeled are grouped into a single “Other” category for clarity.

SHAP scores can be assigned directly to the molecular graph by pairing the predictions of our graph-based MPNN model with the interpretability procedure developed by Li et al.[39] This enables deeper insight into the structural features that the ML model associates with thermal stability and complements the prior analysis of higher-level physical descriptors covered by the RF model. Figure 7 illustrates a selection of targeted structural modifications to TATB and describes how specific atoms and bonds contribute to the relative thermal stability of the resulting molecules.

Inspection of Figure 7a reveals that in TATB the aromatic core and $NH_2$ groups contribute strongly to the total thermal stability of the molecule while the $NO_2$ groups tend to detract from it. This result is in close agreement with the BDE data presented in Figure 6 that shows R-$NH_2$ bonds as being some of the strongest in the dataset while R-$NO_2$ bonds tend to be the weakest. Furthermore, the symmetrical positions of $NH_2$ proton donors and $NO_2$ proton acceptors enable strong inter- and intramolecular H-bonding that acts as a stabilizing force known to contribute to the excellent thermal stability of TATB.[36] Given that TATB already exhibits a high decomposition temperature (approximately 350 °C),[40,41] there are few modifications one can apply to increase it without completely changing the core. However, some chemically intuitive changes such as isolating the aromatic core entirely (Figure 7b) and replacing the energetic $NO_2$ groups with highly stable aromatic rings (Figure 7c) are attempted. In both cases, the model predicts an increase in the $T_d$ by more than 50 °C relative to unmodified TATB. While these specific modifications may result in counterproductive changes to other properties of interest, we stress that this model can be used in conjunction with surrogate models that predict other properties[18,42–45] to drive multi-objective molecule design.

The second row of Figure 7 depicts modifications that are detrimental to predicted $T_d$ and establish design principles for the structural features that should be avoided in the development of new thermally stable energetic materials. In Figure 7d, the stabilizing $NH_2$ groups are replaced with an alternative proton donor, OH. This results in a significant reduction in predicted thermal stability relative to TATB ($\Delta T_d$ = -226 °C). While Figure 6 does show that R-OH bonds tend to be weaker than R-$NH_2$ bonds on average, such a significant drop in decomposition temperature may be an indication that the OH groups are less effective in forming strong stabilizing H-bond networks than the $NH_2$ groups present in unmodified TATB. This is despite the fact that OH groups are stronger H-bond donors than $NH_2$ groups[46] - suggesting that the presence of two available H-bonding sites in each $NH_2$ group may be contributing to a more stabilizing network effect even though each individual interaction is weaker. Additionally, the acidic strength of the OH group relative to $NH_2$ may also contribute to this effect. Figure 7e focuses specifically on weakening the R-$NO_2$ linkage by replacing it with a R-O-$NO_2$ linkage. This results in an even more significant drop in $T_d$ ($\Delta T_d$ = -317 °C) indicating that the new O-$NO_2$ group cannot be as effectively stabilized by the presence of a neighboring proton donor such as $NH_2$. Lastly, Figure 7f modifies the aromatic core through nitrogen heteroatom substitution. The ring nitrogen withdraws electron density from the aromatic framework, altering the electronic environment of the adjacent $NH_2$ group and reducing its ability to participate in stabilizing intramolecular interactions with the neighboring $NO_2$ substituent. This behavior is reflected in the lower SHAP score assigned to the $NH_2$ group bound to the aza-substituted position relative to the $NH_2$ groups bound to unmodified sites. Consequently, disruption of the $NH_2$-$NO_2$ stabilizing interaction produces a decrease in predicted decomposition temperature ($\Delta T_d$ -160 °C). However, this effect size is the smallest of the three destabilizing mechanisms considered.

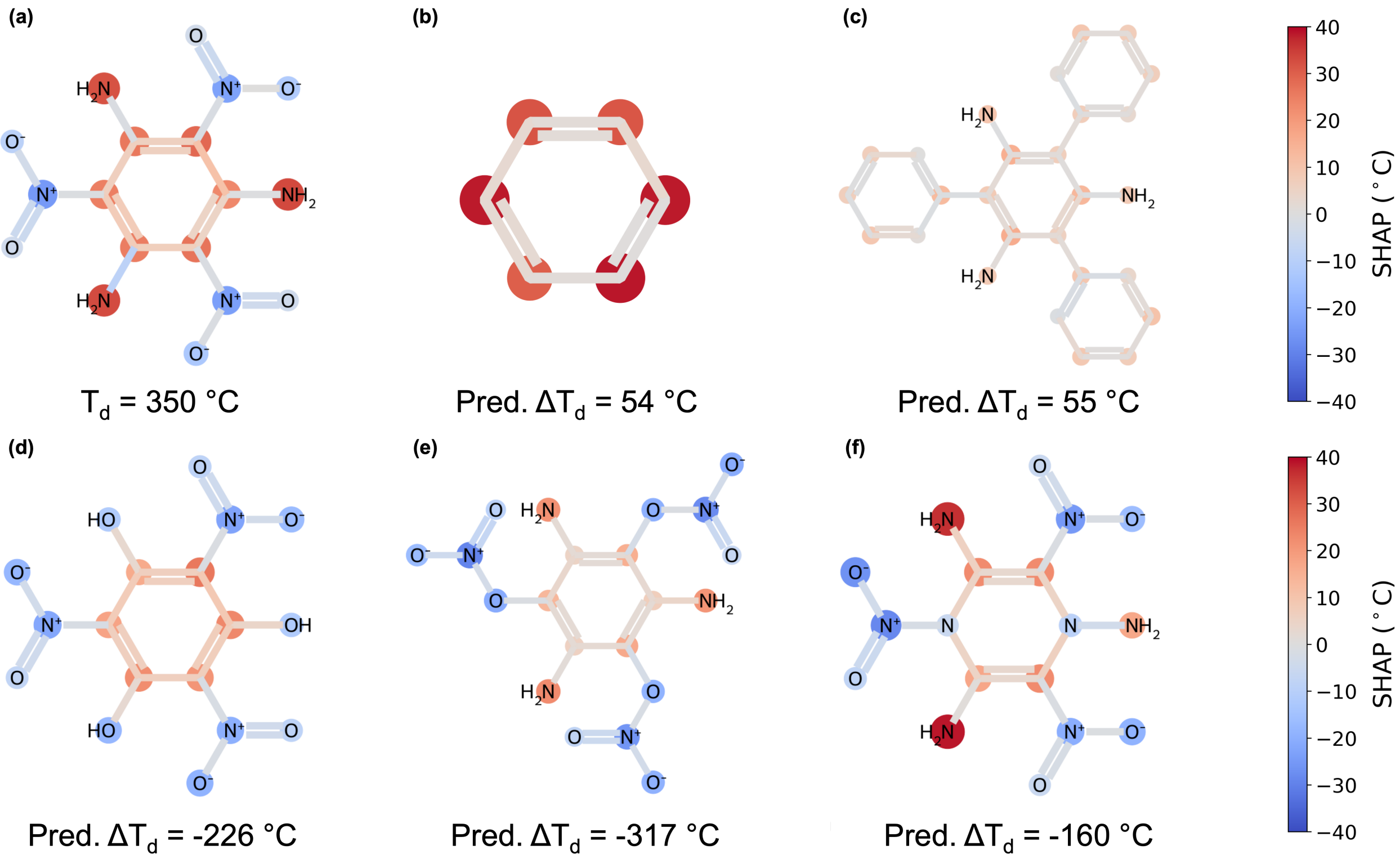

Figure 7 – Molecular graph-based SHAP analysis derived from the predictions of the MPNN model. TATB (a) is chosen as the anchor molecule for all predictions. Subplots (b) and (c) illustrate modifications that are predicted to result in molecules with increased thermal stability relative to TATB by isolating the aromatic core (b) and replacing $NO_2$ groups with aromatic rings (c). Subplots (d), (e), and (f) illustrate modifications that are predicted to result in molecules with reduced thermal stability relative to TATB by replacing $NH_2$ groups with alternative proton donors (d), weakening the R-$NO_2$ bonds (e), and disrupting the aromaticity of the core by introducing heteroatoms (f). Predicted $\Delta T_d$ values relative to TATB are shown as labels below each subplot. Bonds and atoms which contribute positively to the thermal stability are colored red while bonds and atoms that detract from thermal stability are colored blue as indicated by the color bars on the right. Slight variations in the SHAP values of symmetrical positions are expected due to the non-exhaustive nature of the SHAP analysis procedure.

**Conclusions**

In this work, we develop two distinct differential regression models that can be used to rank molecules by their thermal stability. A message-passing neural network that learns directly from the graph representation of a molecule and a random forest model trained on physically interpretable topological and DFT-derived descriptors are shown to have similar predictive performance and exhibit robust ranking metrics. We find that this differential learning strategy outperforms conventional regression methods and is especially useful for data-constrained problems such as thermal stability prediction due to the fact that the number of possible training pairs far exceeds the number of individual molecules. We also highlight how the predictions from the regression models can be post-processed to compute classification metrics that further improve ranking accuracy. To our knowledge, these models exhibit state-of-the-art performance in thermal stability ranking (pairwise accuracy of 87% in the test set) making them useful tools for high-throughput materials discovery applications where thermal stability is a key issue - such as in the realm of energetic materials.

Furthermore, feature importance analysis reveals chemical insights that can inform future development of new thermally stable energetic materials. We find that the difference in oxygen balance and bond dissociation enthalpy between a pair of molecules are key factors in the ability of the model to predict the difference in their thermal stability. However, we stress that the contribution of any individual descriptor is small relative to the sum of all descriptors, which indicates that within the set of molecular descriptors examined in this study, no individual property should be used directly as a proxy for thermal stability. Instead, more complex non-linear relationships between the descriptors are what enable the ML models to make accurate predictions. Taking TATB as a case study, we also apply graph-level SHAP analysis to highlight the importance of stabilizing H-bond networks as a key structural feature to guide the development of new thermally stable materials.

Lastly, we make the complete dataset publicly available to facilitate future model development and benchmarking. An important next step will be the construction of a decomposition temperature database that includes standardized metadata describing the experimental conditions associated with each measurement. We anticipate that reducing variability arising from differences in testing protocols and reporting practices will improve the quality of the training data and, consequently, the predictive performance of future models. We also note that the molecular descriptors employed in this work are derived primarily from gas-phase representations of isolated molecules. As such, the present models do not explicitly account for condensed-phase effects, including intermolecular interactions, crystal packing, and other collective phenomena that may meaningfully influence thermal stability in real materials. While the strong performance of the current models suggests that intrinsic molecular features capture a significant fraction of the observed behavior, condensed-phase properties likely provide additional predictive information that remains unexplored. Future work will therefore focus on incorporating descriptors derived from calculated and experimental condensed-phase properties to quantify the contribution of intermolecular interactions and further improve model accuracy and interpretability.

## 4. Methods

### *4.1 Workflow and data collection*

We initially collect molecules and their associated experimental decomposition temperatures from the work of Wu et al.[6] and Mathieu et al.[2] The data from Mathieu et al. can be easily extracted directly from their reported supplemental information documents. In Wu et al., the molecules are provided as images of molecular graphs and the associated $T_d$ values are stored in the images as labels. The molecular graph images were converted into SMILES strings using a selection of three machine learning models: DECIMER[47–49], MolNexTR[50], and MolScribe.[51] If any two of the models predicted the same SMILES string, then that molecule was considered valid and added to the dataset. The Python library EasyOCR[52] was used to convert the labels in the images to numerical $T_d$ values. On the occasion that each source reported a different $T_d$ for the same molecule, the average of the two reports was taken to be the target value. This results in a dataset of 915 molecules with experimentally reported decomposition temperatures. 57 molecules were separated out to form a held-out testing set using the binned-sampling strategy outlined in Section 4.3, leaving 858 molecules for training.

### *4.2 Density functional theory calculations*

We produce an initial geometry for each molecule using the Confab conformer generation routine[53] in Open Babel.[54] This method creates structurally diverse low-energy conformers by evaluating the energy barrier to rotate different bonds present within the molecule and comparing the root mean square deviation (RMSD) of different conformers. The routine uses the GFN2-xTB[55] tight-binding method to select the lowest-energy conformer for higher-fidelity DFT calculations. These initial structures are all optimized at the $\omega$B97X-D/6-311G** level of theory[56] as implemented in Gaussian 16 (vC.01).[57] Following DFT optimization, we identify all possible homolytic bond dissociation reactions for non-ring bonds in each molecule using RDKit[58] and apply them to produce radical fragments $f_1$ and $f_2$. The initial geometry of the radical fragments is set to correspond to the coordinates of the corresponding fragment in the optimized parent molecule. The radical fragments are then optimized using the same level of theory used to optimize the parents. Bond dissociation enthalpy is computed with the following equation:

$$BDE = (H_{f1} + H_{f2}) - H_{mol} \qquad (1)$$

Where $H$ represents the total enthalpy of the parent molecule ($H_{mol}$) and each radical fragment ($H_{f1}$, $H_{f2}$). Note that in rare cases the fragmentation procedure may produce a negative BDE due to especially low energy products as compared to the parent molecule. In addition to the enthalpy used to compute BDEs, a list of additional DFT-computed properties collected to be used as ML descriptors is provided in Supplemental Information Table S2. All DFT job submission and data collection is managed with the Python framework *pyiron.*[59]

*4.3 Machine learning methods*

We develop a message-passing neural network (MPNN) implemented using the Python package Chemprop[35] as a surrogate model to predict the difference in decomposition temperature ($\Delta T_d$) between pairs of molecules. Models with this architecture represent atoms and bonds as nodes and edges of a molecular graph. Information is exchanged between bonded atoms via message-passing to update the model weights during training. After multiple rounds of message-passing, atom and bond-level features are pooled into a single molecule-level embedding that is passed to a feed-forward neural network, which predicts the target properties. While this architecture is typically applied to individual molecules, it is straightforward to extended it to pairs of molecules, as we do in this work, by treating the two input molecules as a single graph with disconnected components. This treatment is supported directly in Chemprop by the *MulticomponentDataset* class.

Molecular graphs are read as SMILES strings, canonicalized using RDKit, and filtered to remove duplicate or malformed entries. A unique set of 10,000 pairs are generated from the molecules in the training set according to a stratified sampling procedure in which molecules were grouped by their $T_d$ into ten bins and chosen such that each bin has nearly equal representation in the final subset. Selecting molecules in this way enforces a balanced representation of molecular pairs across the entire range of $\Delta T_d$ values and enables the model to effectively learn from regions that would otherwise be underrepresented in the dataset. This procedure is repeated 10 times to produce unique splits for cross-fold validation. Targets are standardized to the mean and variance of the training set, and an inverse-transform layer unscales the predictions at inference. Pairs of molecules are featurized with the default *SimpleMoleculeMolGraphFeaturizer* class available in Chemprop. A dropout layer with a rate of 20% is applied to the bond message-passing and final feed-forward steps to prevent overfitting and batch normalization is applied after aggregation to stabilize training. Hyperparameter optimization is managed by the Python package Ray Tune,[60] which employs the Asynchronous Successive Halving Algorithm (ASHA)[61] for efficient early-stopping. The optimized hyperparameters are outlined in Supplemental Information Table S3. Each trial is trained for up to 100 epochs with ASHA configured to use a grace period of 20 epochs and a reduction factor of 2. Hinge loss, $L = \max(0, 1 - t \cdot y)$, (where $t$ is the ground-truth and $y$ is the predicted value) is chosen as the loss function. This loss function only penalizes the model when the sign of the $\Delta T_d$ prediction is incorrect; resulting in a model that prioritizes accuracy in the pairwise ordering of molecules more so than minimizing the magnitude of the $\Delta T_d$ prediction error of individual pairs.[62] Adam[63] is used as the optimizer in conjunction with a Noam learning-rate scheduler,[64] which scales the learning rate from an initial value of $1 \times 10^{-4}$ to a final value of $1 \times 10^{-5}$. Model checkpoints are saved for each trial, and the configuration that produces the lowest validation error in each fold is retained for inference. All training is conducted on NVIDIA A100 Tensor Core GPUs, with four GPUs per node used for parallel execution.

In addition to the graph-based models outlined above, descriptor-based models are also trained with a differential learning strategy. These models are implemented using scikit-learn 1.6.1.[34] Differential regression was found to outperform classification (see Table S1). Several tree-based models were tested, including RandomForestRegressor, GradientBoostingRegressor and ExtraTreesRegressor. We attempted to implement hinge-loss criterion for split impurity for trees, but this was outperformed by standard regression criterion (Table S1). The input representation for these models is constructed from DFT descriptors computed from our BDE workflow together with RDKit cheminformatic descriptors (including features dependent on optimized 3D geometry from DFT) and several additional hand-crafted features. The latter includes heats of sublimation computed via the group-additive model reported by Mathieu,[65] force constants for minimum BDE bonds computed with the LMODEA package,[66] the bond-type of the minimum BDE (one-hot encoded), and density, which is estimated by using RDKit to compute the molecular volume of the lowest energy MMFF94[67] conformer then fitting to a dataset of experimental density measurements as we have described in prior work.[45] Heat of explosion ($H_{exp}$), which describes the energy released as a molecule is converted into its primitive detonation products, is computed as the difference between the solid-state heat of formation of the reactant molecule and the sum of the reference energies of the following products weighted by their stoichiometric coefficients: $N_2$, $H_2O$, $CO_2$, $O_2$, C (graphite), CO, and $H_2$. The amount of each product is determined by the Modified Kistiakowsky-Wilson rules.[68] The reference energies are computed via DFT using the same level of theory described in Section 4.2. Lastly, oxygen balance ($\%OB$) is calculated according to the following equation,

$$\%OB = \frac{-1600(2N_C+0.5N_H-N_O)}{M} \qquad (2)$$

where $N_C$, $N_H$, and $N_O$ represent the number of carbon, hydrogen, and oxygen atoms respectively, and $M$ is the molecular weight. Using Equation 2 to compute the oxygen balance results in all sources of oxygen in the molecule counting towards the final value. This is conventional, but differs from another formulation by Kamlet and Adolph[69] which excludes oxygen atoms in carbonyl groups.

The full list of descriptors is given in Table S2. Three approaches for input representation are examined: (i) using descriptor matrices for molecule *A* and molecule *B* concatenated together, (ii) using the difference of the input vectors for molecule *A* and *B* (differential descriptors), and (iii) both approaches combined. It is found that the second approach, differential descriptors, has the best performance (Table S1) which is consistent with prior work.[70] Furthermore, it is the representation most amenable to SHAP[71,72] analysis. The input vectors are then pruned by removing descriptors with zero variance and with > 0.95 Pearson correlation with any other descriptor. The list of pruned descriptors is given in Supplemental Information Table S4.

10,000 pairs for training are sampled from the training set in the same manner (ensuring an even distribution across the range of $T_d$ values) as for the MPNN model. Hyperparameter optimization is performed by constructing 20 cross-validation folds with a 90:10 train:validation split and using HalvingGridSearchCV to find optimal hyperparameters for each model (hyperpameter grids are given in Supplemental Information Table S5). Once hyperparameters are selected, a BaggingRegressor meta-estimator is constructed for each model, forming an ensemble of 20 models trained on separate datasets constructed via sampling-with-replacement from the training data. Predictions from each sub-estimator are then aggregated to provide final predictions and standard deviations for predictions and performance metrics. The SHAP package is used for feature importance analysis.

### *4.4 Classification and ranking metrics*

The MPNN and RF models are both regressors trained to predict a scalar $\Delta T_d$ for a given pair of molecules. However, through postprocessing, we convert these regression results into a binary classification problem indicating whether molecule $A$ or molecule $B$ is predicted to have a higher $T_d$. If the sign of $\Delta T_d$ is negative, the binary prediction is 0 and if the sign of $\Delta T_d$ is positive the binary prediction is 1. Transforming the problem in this way is especially useful for noisy datasets such as ours because the noise in the continuous target variable leads to large regression residuals that obscure meaningful structure whereas classification of similarly noisy datasets tends to be more robust.[73,74] Given that our goal is to guide the development of materials with improved thermal stability relative to the current state of the art, preserving the ordinality of the pairwise predictions is more valuable than precisely predicting the difference between two molecules.

To support this goal, we develop a simple tournament-based ranking system to score molecules and assign them a rank based on their implied thermal stability. All possible pairs of molecules in the test set are constructed and evaluated with the surrogate model to generate a $\Delta T_d$ prediction for each pair. For each individual molecule, its corresponding pairs are analyzed and the number of pairs in which that molecule was predicted to have a higher $T_d$ (a binary classification of 1) are summed together to produce a "win count" metric. Each molecule is then ranked by its win count such that the molecule with the most wins is interpreted as being the most thermally stable. To assess the quality of a predicted ranking we compute pairwise accuracy, a measure of the percentage of samples that are correctly ordered relative to their ground-truth rank, according to the following equation,

$$A_{pair} = \frac{(1+\tau)}{2} \qquad (3)$$

where $\tau$ is Kendall's tau[75] evaluated between the ground-truth ordering and the predicted ordering. While this system is effective and robust, it suffers from $O(N^2)$ scaling. Fortunately, this can be reduced to $O(N)$ if a single "anchor molecule" with a known $T_d$ is compared against all unknown molecules. In this anchored system, a precise $T_d$ can be derived for each unknown molecule by simply adding the $T_d$ of the anchor molecule to the $\Delta T_d$ predicted by the model. Molecules can then be ranked directly by their derived $T_d$ without resorting to the more computationally expensive win count metric. While this anchor-based system is faster for large datasets, in this work we choose to rely on the tournament-based system as we expect that comparison against multiple molecules will result in a more robust predicted ordering.

**Corresponding Authors**

megand@lanl.gov, sullberg@lanl.gov, kortkamp@lanl.gov

**Author Contributions**

WKK, IM, and CJS conceived the project. MCD, RSU, IM, and JNS developed the high-throughput workflows to run density functional theory calculations. JNS calculated the DFT and BDE descriptors. MCD curated the cheminformatics descriptors, trained the tree-based regression models, and ran the SHAP analysis of the RF model. RSU curated the molecules from literature, ran DFT calculations, and ran the SHAP analysis of the MPNN model. RSU and WKK trained the MPNN models. MCD and RSU wrote the manuscript with input from WKK, IM, MJC, and CJS. All authors have approved of the final version of the manuscript. MCD and RSU contributed equally. WKK and IM supervised the execution of the project.

## Data Availability

A dataset of all the SMILES strings, $T_d$ targets, and descriptor values used in this work is available for download at: https://zenodo.org/records/22033730.

## Acknowledgements

Research presented in this article was supported by the Laboratory Directed Research and Development program of Los Alamos National Laboratory under project number 20250006DR. The authors thank the NNSA Minority Serving Institution Partnership Program (MSIPP) for additional financial support. This research used resources provided by the Los Alamos National Laboratory Institutional Computing Program which is supported by the U.S. Department of Energy National Nuclear Security Administration under contract No. 89233218C-NA000001. This work has been approved for unlimited release under LA-UR-26-27497.

Supplemental Information for the Manuscript:

# Differential Learning for Robust Prediction of Thermal Stability with Application to Energetic Materials

*Megan C. Davis*[†*], *R. Seaton Ullberg*[†*], *Jeremy N. Schroeder*[†‡], *Andrew H. Salij* [†], *Marc J. Cawkwell*[†], *Christopher J. Snyder*[¶], *Ivana Matanovic*[†], *Wilton J. M. Kort-Kamp*[†]

† Theoretical Division, Los Alamos National Laboratory, Los Alamos, NM 87545, United States

‡ Department of Mechanical and Aerospace Engineering, Texas Tech University, Lubbock, TX 79409, United States

¶ Weapon Stockpile Modernization Division, Los Alamos National Laboratory, Los Alamos, NM 87545, United States

Corresponding authors: megand@lanl.gov, sullberg@lanl.gov, kortkamp@lanl.gov

[*] These authors contributed equally.

**Table of Contents**

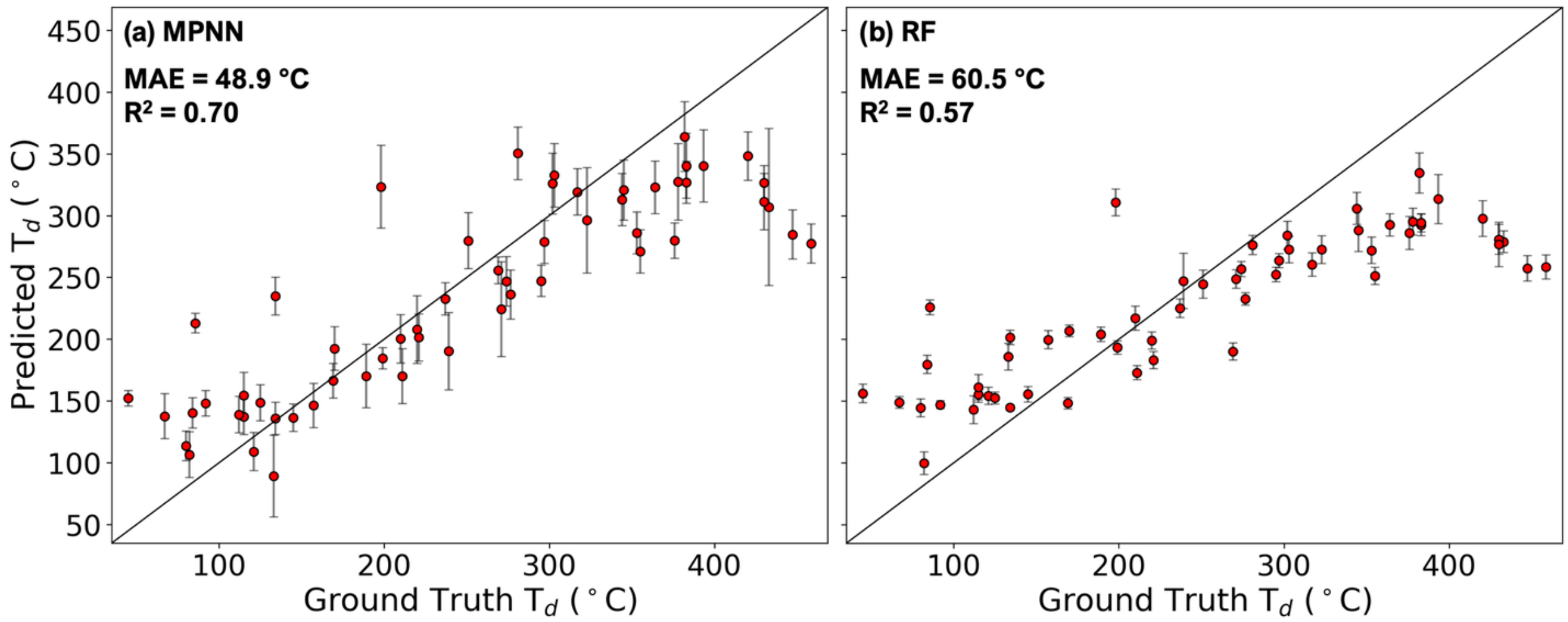

Figure S1 – Parity plots illustrating the predictive performance of direct regression models for (a) the MPNN architecture and (b) the RF architecture. High MAE and low $R^2$ motivate the decision to move forward with a differential regression strategy.

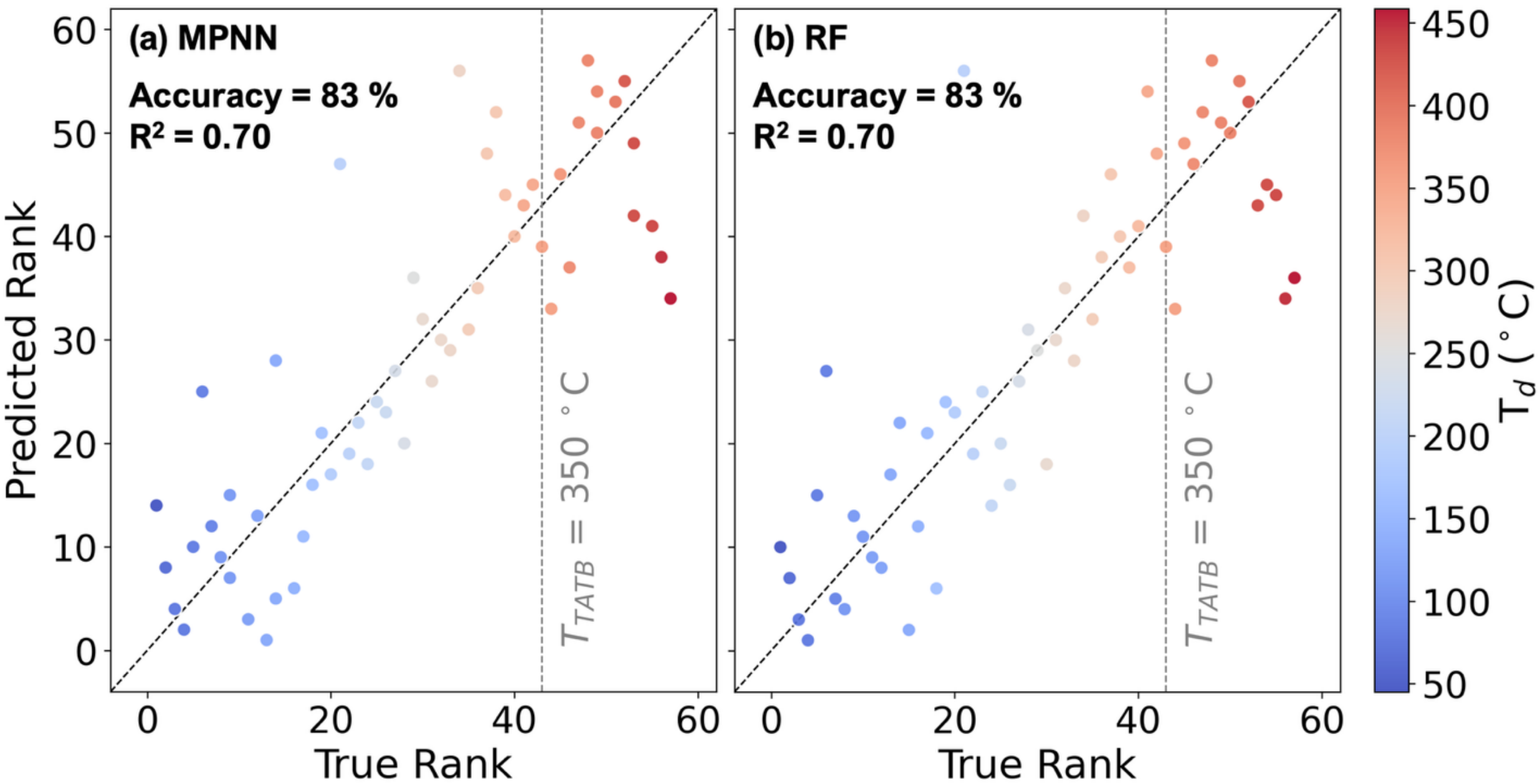

Figure S2 – Parity of (a) the direct regression MPNN model and (b) the direct regression RF model in predicting the order of each molecule in the test set ranked by its $T_d$. Points are colored by their $T_d$ with red corresponding to higher values and blue corresponding to lower values as indicated by the color bar on the right. The position of TATB is indicated by a dashed line in both subplots to give a sense of the $T_d$ range which the test molecules cover.

Table S1 - Model Performance of descriptor based differential models for different input representations and model architectures. “Delta” refers to differential descriptors defined in the main text. “Combined” refers to concatenating descriptors for both input molecules. “Both” refers to concatenating “Delta” and “Combined” descriptors. “Hinge-Delta-GBT” and “Hinge-Delta-RF” are models where a hinge loss function was used as the tree split criterion for regression trees. “Reg” indicates that the model is a regression model, otherwise it is a classifier.

| **Model** | **F1** | **Accuracy** | **AUROC** | **AUPRC** |
|---|---|---|---|---|
| Delta-RF | 0.845 +/- 0.011 | 0.845 +/- 0.013 | 0.923 +/- 0.004 | 0.928 +/- 0.004 |
| Delta-XTR | 0.835 +/- 0.010 | 0.833 +/- 0.011 | 0.916 +/- 0.005 | 0.922 +/- 0.004 |
| Delta-GBT | 0.830 +/- 0.011 | 0.826 +/- 0.012 | 0.916 +/- 0.008 | 0.922 +/- 0.007 |
| Both-RF | 0.837 +/- 0.005 | 0.835 +/- 0.005 | 0.921 +/- 0.003 | 0.924 +/- 0.003 |
| Both-XTR | 0.840 +/- 0.01 | 0.838 +/- 0.008 | 0.919 +/- 0.005 | 0.923 +/- 0.005 |
| Both-GBT | 0.833 +/- 0.014 | 0.831 +/- 0.014 | 0.926 +/- 0.005 | 0.930 +/- 0.005 |
| Combined-RF | 0.828 +/- 0.009 | 0.826 +/- 0.009 | 0.906 +/- 0.005 | 0.907 +/- 0.007 |
| Combined-XTR | 0.835 +/- 0.010 | 0.833 +/- 0.010 | 0.915 +/- 0.005 | 0.919 +/- 0.006 |
| Combined-GBT | 0.826 +/- 0.014 | 0.824 +/- 0.015 | 0.920 +/- 0.008 | 0.925 +/- 0.008 |
| Hinge-Delta-GBT | 0.852 +/- 0.008 | 0.850 +/- 0.008 | 0.934 +/- 0.004 | 0.939 +/- 0.004 |
| Hinge-Delta-RF | 0.811 +/- 0.021 | 0.813 +/- 0.021 | 0.889 +/- 0.010 | 0.896 +/- 0.009 |
| **Reg-Delta-RF** | **0.860 +/- 0.007** | **0.861 +/- 0.007** | **0.936 +/- 0.002** | **0.940 +/- 0.002** |
| Reg-Delta-XTR | 0.853 +/- 0.008 | 0.854 +/- 0.008 | 0.929 +/- 0.004 | 0.934 +/- 0.003 |
| Reg-Delta-GBT | 0.845 +/- 0.010 | 0.842 +/- 0.010 | 0.932 +/- 0.005 | 0.937 +/- 0.004 |
| Reg-Both-RF | 0.838 +/- 0.008 | 0.835 +/- 0.009 | 0.928 +/- 0.003 | 0.933 +/- 0.002 |
| Reg-Both-XTR | 0.842 +/- 0.008 | 0.840 +/- 0.009 | 0.929 +/- 0.003 | 0.933 +/- 0.003/ |
| Reg-Both-GBT | 0.833 +/- 0.017 | 0.832 +/- 0.018 | 0.923 +/- 0.010 | 0.926 +/- 0.010 |
| Reg-Combined-RF | 0.837 +/- 0.010 | 0.834 +/- 0.010 | 0.919 +/- 0.004 | 0.921 +/- 0.004 |
| Reg-Combined-XTR | 0.833 +/- 0.007 | 0.830 +/- 0.007 | 0.923 +/- 0.003 | 0.928 +/- 0.003 |
| Reg-Combined-GBT | 0.835 +/- 0.009 | 0.832 +/- 0.009 | 0.923 +/- 0.006 | 0.927 +/- 0.006 |

Table S2 - All descriptors considered for descriptor-based ML models together with source: DFT calculations (DFT), RDKit (RDKit) or uniquely crafted descriptors (Custom).

| Descriptor | Source |
|---|---|
| A | DFT |
| B | DFT |
| C | DFT |
| mu | DFT |
| alpha | DFT |
| alpha_homo | DFT |
| alpha_lumo | DFT |
| gap | DFT |
| r2 | DFT |
| U0 | DFT |
| U | DFT |
| H (kcal/mol) | DFT |
| G | DFT |
| Cv | DFT |
| dipole_x | DFT |
| dipole_y | DFT |
| dipole_z | DFT |
| quad_xx | DFT |
| quad_yy | DFT |
| quad_zz | DFT |
| quad_xy | DFT |

| quad_xz | DFT |
|---|---|
| quad_yz | DFT |
| max_charge | DFT |
| min_charge | DFT |
| Hf_gas (kcal/mol) | DFT |
| Hf_solid (kcal/mol) | DFT |
| Hexp (kcal/mol) | DFT |
| BDE_min_fc (mDyn/A) | DFT |
| BDE_min_vf (cm^-1) | DFT |
| BDE_min_bt | DFT |
| BDE_min (kcal/mol) | DFT |
| BDE_mean (kcal/mol) | DFT |
| Hsub | Custom |
| PMI1 | RDKit |
| PMI2 | RDKit |
| PMI3 | RDKit |
| NPR1 | RDKit |
| NPR2 | RDKit |
| RadiusOfGyration | RDKit |
| InertialShapeFactor | RDKit |
| Eccentricity | RDKit |
| Asphericity | RDKit |
| SpherocityIndex | RDKit |
| PBF | RDKit |
| MaxAbsEStateIndex | RDKit |
| MaxEStateIndex | RDKit |
| MinAbsEStateIndex | RDKit |
| MinEStateIndex | RDKit |
| qed | RDKit |
| SPS | RDKit |
| MolWt | RDKit |
| HeavyAtomMolWt | RDKit |
| ExactMolWt | RDKit |
| NumValenceElectrons | RDKit |
| NumRadicalElectrons | RDKit |
| MaxPartialCharge | RDKit |
| MinPartialCharge | RDKit |
| MaxAbsPartialCharge | RDKit |
| MinAbsPartialCharge | RDKit |
| FpDensityMorgan1 | RDKit |
| FpDensityMorgan2 | RDKit |
| FpDensityMorgan3 | RDKit |
| BCUT2D_MWHI | RDKit |
| BCUT2D_MWLOW | RDKit |
| BCUT2D_CHGHI | RDKit |
| BCUT2D_CHGLO | RDKit |
| BCUT2D_LOGPHI | RDKit |
| BCUT2D_LOGPLOW | RDKit |
| BCUT2D_MRHI | RDKit |
| BCUT2D_MRLOW | RDKit |
| AvgIpc | RDKit |
| BalabanJ | RDKit |
| BertzCT | RDKit |
| Chi0 | RDKit |
| Chi0n | RDKit |
| Chi0v | RDKit |
| Chi1 | RDKit |
| Chi1n | RDKit |
| Chi1v | RDKit |
| Chi2n | RDKit |
| Chi2v | RDKit |
| Chi3n | RDKit |
| Chi3v | RDKit |
| Chi4n | RDKit |
| Chi4v | RDKit |
| HallKierAlpha | RDKit |
| Ipc | RDKit |
| Kappa1 | RDKit |
| Kappa2 | RDKit |
| Kappa3 | RDKit |
| LabuteASA | RDKit |
| PEOE_VSA1 | RDKit |
| PEOE_VSA10 | RDKit |
| PEOE_VSA11 | RDKit |
| PEOE_VSA12 | RDKit |
| PEOE_VSA13 | RDKit |
| PEOE_VSA14 | RDKit |
| PEOE_VSA2 | RDKit |
| PEOE_VSA3 | RDKit |
| PEOE_VSA4 | RDKit |
| PEOE_VSA5 | RDKit |
| PEOE_VSA6 | RDKit |
| PEOE_VSA7 | RDKit |
| PEOE_VSA8 | RDKit |
| PEOE_VSA9 | RDKit |
| SMR_VSA1 | RDKit |
| SMR_VSA10 | RDKit |
| SMR_VSA2 | RDKit |
| SMR_VSA3 | RDKit |
| SMR_VSA4 | RDKit |
| SMR_VSA5 | RDKit |
| SMR_VSA6 | RDKit |
| SMR_VSA7 | RDKit |
| SMR_VSA8 | RDKit |
| SMR_VSA9 | RDKit |
| SlogP_VSA1 | RDKit |
| SlogP_VSA10 | RDKit |
| SlogP_VSA11 | RDKit |
| SlogP_VSA12 | RDKit |
| SlogP_VSA2 | RDKit |
| SlogP_VSA3 | RDKit |
| SlogP_VSA4 | RDKit |
| SlogP_VSA5 | RDKit |
| SlogP_VSA6 | RDKit |
| SlogP_VSA7 | RDKit |
| SlogP_VSA8 | RDKit |
| SlogP_VSA9 | RDKit |
| TPSA | RDKit |
| EState_VSA1 | RDKit |
| EState_VSA10 | RDKit |
| EState_VSA11 | RDKit |
| EState_VSA2 | RDKit |
| EState_VSA3 | RDKit |
| EState_VSA4 | RDKit |
| EState_VSA5 | RDKit |
| EState_VSA6 | RDKit |
| EState_VSA7 | RDKit |

| EState_VSA8 | RDKit |
|---|---|
| EState_VSA9 | RDKit |
| VSA_EState1 | RDKit |
| VSA_EState10 | RDKit |
| VSA_EState2 | RDKit |
| VSA_EState3 | RDKit |
| VSA_EState4 | RDKit |
| VSA_EState5 | RDKit |
| VSA_EState6 | RDKit |
| VSA_EState7 | RDKit |
| VSA_EState8 | RDKit |
| VSA_EState9 | RDKit |
| FractionCSP3 | RDKit |
| HeavyAtomCount | RDKit |
| NHOHCount | RDKit |
| NOCount | RDKit |
| NumAliphaticCarbocycles | RDKit |
| NumAliphaticHeterocycles | RDKit |
| NumAliphaticRings | RDKit |
| NumAmideBonds | RDKit |
| NumAromaticCarbocycles | RDKit |
| NumAromaticHeterocycles | RDKit |
| NumAromaticRings | RDKit |
| NumAtomStereoCenters | RDKit |
| NumBridgeheadAtoms | RDKit |
| NumHAcceptors | RDKit |
| NumHDonors | RDKit |
| NumHeteroatoms | RDKit |
| NumHeterocycles | RDKit |
| NumRotatableBonds | RDKit |
| NumSaturatedCarbocycles | RDKit |
| NumSaturatedHeterocycles | RDKit |
| NumSaturatedRings | RDKit |
| NumSpiroAtoms | RDKit |
| NumUnspecifiedAtomStereoCenters | RDKit |
| Phi | RDKit |
| RingCount | RDKit |
| MolLogP | RDKit |
| MolMR | RDKit |

| fr_Al_COO | RDKit |
|---|---|
| fr_Al_OH | RDKit |
| fr_Al_OH_noTert | RDKit |
| fr_ArN | RDKit |
| fr_Ar_COO | RDKit |
| fr_Ar_N | RDKit |
| fr_Ar_NH | RDKit |
| fr_Ar_OH | RDKit |
| fr_COO | RDKit |
| fr_COO2 | RDKit |
| fr_C_O | RDKit |
| fr_C_O_noCOO | RDKit |
| fr_C_S | RDKit |
| fr_HOCCN | RDKit |
| fr_Imine | RDKit |
| fr_NH0 | RDKit |
| fr_NH1 | RDKit |
| fr_NH2 | RDKit |
| fr_N_O | RDKit |
| fr_Ndealkylation1 | RDKit |
| fr_Ndealkylation2 | RDKit |
| fr_Nhpyrrole | RDKit |
| fr_SH | RDKit |
| fr_aldehyde | RDKit |
| fr_alkyl_carbamate | RDKit |
| fr_alkyl_halide | RDKit |
| fr_allylic_oxid | RDKit |
| fr_amide | RDKit |
| fr_amidine | RDKit |
| fr_aniline | RDKit |
| fr_aryl_methyl | RDKit |
| fr_azide | RDKit |
| fr_azo | RDKit |
| fr_barbitur | RDKit |
| fr_benzene | RDKit |
| fr_benzodiazepine | RDKit |
| fr_bicyclic | RDKit |
| fr_diazo | RDKit |
| fr_dihydropyridine | RDKit |

| fr_epoxide | RDKit |
|---|---|
| fr_ester | RDKit |
| fr_ether | RDKit |
| fr_furan | RDKit |
| fr_guanido | RDKit |
| fr_halogen | RDKit |
| fr_hdrzine | RDKit |
| fr_hdrzone | RDKit |
| fr_imidazole | RDKit |
| fr_imide | RDKit |
| fr_isocyan | RDKit |
| fr_isothiocyan | RDKit |
| fr_ketone | RDKit |
| fr_ketone_Topliss | RDKit |
| fr_lactam | RDKit |
| fr_lactone | RDKit |
| fr_methoxy | RDKit |
| fr_morpholine | RDKit |
| fr_nitrile | RDKit |
| fr_nitro | RDKit |
| fr_nitro_arom | RDKit |
| fr_nitro_arom_nonortho | RDKit |
| fr_nitroso | RDKit |
| fr_oxazole | RDKit |
| fr_oxime | RDKit |
| fr_para_hydroxylation | RDKit |
| fr_phenol | RDKit |
| fr_phenol_noOrthoHbond | RDKit |
| fr_phos_acid | RDKit |
| fr_phos_ester | RDKit |
| fr_piperdine | RDKit |
| fr_piperzine | RDKit |
| fr_priamide | RDKit |
| fr_prisulfonamd | RDKit |
| fr_pyridine | RDKit |
| fr_quatN | RDKit |
| fr_sulfide | RDKit |
| fr_sulfonamd | RDKit |
| fr_sulfone | RDKit |

| fr_term_acetylene | RDKit |
|---|---|
| fr_tetrazole | RDKit |
| fr_thiazole | RDKit |
| fr_thiocyan | RDKit |
| fr_thiophene | RDKit |
| fr_unbrch_alkane | RDKit |
| fr_urea | RDKit |
| O_balance | Custom |

Table S3 – Optimized Chemprop model hyperparameters.

| **Hyperparameter** | **Optimal Value** |
|---|---|
| Message Passing | |
| d_e | 14 |
| d_h | 950 |
| d_v | 72 |
| d_vd | None |
| depth | 5 |
| dropout | 0.2 |
| Feed Forward | |
| hidden_dim | 350 |
| input_dim | 950 |
| dropout | 0.2 |

Table S4 - Descriptors (Kept Descriptor) which had >0.95 Pearson correlation coefficient with another descriptor (Correlated Descriptors). The latter were removed from the input representation prior to model training.

| **Kept Descriptor** | **Correlated Descriptors** |
|---|---|
| B | C |
| alpha | Chi0n, Chi0v, Chi1n, Chi1v, MolMR |
| r2 | PMI2, PMI3 |

| U0 | U, H (kcal/mol), G, quad_yy, quad_zz |
|---|---|
| Cv | MolWt, HeavyAtomMolWt, ExactMolWt, NumValenceElectrons, Chi0, Chi0n, Chi0v, Chi1n, Chi1v, Kappa1, LabuteASA, HeavyAtomCount |
| Hf_gas (kcal/mol) | Hf_solid (kcal/mol) |
| MaxAbsEStateIndex | MaxEStateIndex |
| FpDensityMorgan1 | FpDensityMorgan2 |
| Kappa2 | Phi |
| SMR_VSA2 | fr_nitrile |
| SMR_VSA3 | fr_Ar_N |
| SlogP_VSA4 | EState_VSA10 |
| TPSA | NOCount, NumHeteroatoms |
| NumAmideBonds | fr_amide |
| NumAromaticCarbocycles | fr_benzene |
| NumAtomStereoCenters | NumUnspecifiedAtomStereoCenters |
| fr_Ar_NH | fr_Nhpyrrole |
| fr_Ar_OH | fr_phenol, fr_phenol_noOrthoHbond |
| fr_COO | fr_COO2 |
| fr_Ndealkylation2 | fr_piperdine |

Table S5 - Hyperparameter grid for best RF model discussed in main text. Optimized hyperparameters are bolded. All other parameters were kept as scikit-learn defaults.

| Parameter | Values |
|---|---|
| n_estimators | 600, **1000** |
| max_depth | **None**, 30 |
| criterion | **squared_error**, friedman_mse |
| min_samples_split | **2**, 10 |
| min_samples_leaf | **1**, 2, 5 |
| max_features | **sqrt**, 0.5 |
| bootstrap | **False**, True |

| min_impurity_decrease | **0.0** |
|---|---|